\documentclass[3p,times]{elsarticle}

\usepackage{ecrc}
\usepackage{gensymb}                   
\usepackage{bbold}
\usepackage{graphicx}
\usepackage{amsmath}
\usepackage{textcomp}
\usepackage{dcolumn}
\usepackage{bm}
\usepackage{amssymb}
\usepackage{amsfonts}
\usepackage{lineno}
\usepackage{url}
\usepackage{array}
\usepackage{verbatim}
\usepackage{subfigure}
\usepackage{listings}
\usepackage{color}
\usepackage{lscape}
\usepackage{setspace}
\usepackage{footmisc}
\usepackage{wasysym}
\usepackage [english]{babel}
\usepackage [autostyle, english = american]{csquotes}

\MakeOuterQuote{"}
\modulolinenumbers[5]

\volume{00}
\firstpage{1}
\journalname{Cryogenics}
\runauth{F.M. Piegsa et al.}

\begin{document}
\begin{frontmatter}

\title{High-power closed-cycle $^4$He cryostat with top-loading sample exchange}

\author[IPP,UBE]{F.M. Piegsa\corref{cor1}}
\ead{florian.piegsa@lhep.unibe.ch}
\author[PSI]{B. van den Brandt}
\author[IPP,PSI]{K. Kirch}

\cortext[cor1]{Corresponding author.}

\address[IPP] {Institute for Particle Physics, ETH Z\"urich, CH-8093 Z\"urich, Switzerland}
\address[UBE] {Laboratory for High Energy Physics, Albert Einstein Center for Fundamental Physics, University of Bern, CH-3012 Bern, Switzerland}
\address[PSI] {Paul Scherrer Institute, CH-5232 Villigen PSI, Switzerland}

\begin{abstract}
We report on the development of a versatile cryogen-free laboratory cryostat based upon a commercial pulse tube cryocooler. It provides enough cooling power for continuous recondensation of circulating $^4$He gas at a condensation pressure of approximately 250~mbar. 
Moreover, the cryostat allows for exchange of different cryostat-inserts as well as fast and easy "wet" top-loading of samples directly into the 1 K pot with a turn-over time of less than 75~min. Starting from room temperature and using a $^4$He cryostat-insert, a base temperature of 1.0~K is reached within approximately seven hours and a cooling power of 250~mW is established at 1.24~K. 

\end{abstract}


\end{frontmatter}

\twocolumn

\section{Introduction}
\label{sec:introduction}
In recent years significant progress in the field of dry refrigerator systems has been made, for instance by Godfrin \cite{Prouve2007,Prouve2008}, Kirichek \cite{Kirichek2011146,Kirichek2013}, Uhlig \cite{Uhlig200273,Uhlig2008511,Uhlig20156}, and Wang \cite{Wang2005719,Wang20145} together with their respective co-workers. 
In the meantime, some of these and similar systems have become commercially available and are now in use in many laboratories employing low temperatures world-wide.
Dry refrigerators usually use Gifford-McMahon or pulse tube cryocoolers (PTC) to liquefy helium gas which is then used in dilution or evaporation stages in order to reach temperatures beyond the intrinsic technical reach of cryocoolers, i.e.\ below 2.5~K.\\ 
This article describes the design and test measurement results of a high-power dry closed-cycle $^4$He evaporation cryostat using a PTC. It offers exchange of different cryostat-inserts as well as top-loading sample exchange. The design of the insert is based on an early development of so-called Roubeau-type cryostats \cite{Roubeau1976,Roubeau1978} which has later been refined and used in several particle physics experiments at the Paul Scherrer Insititute \cite{vandenBrandt1990526}.
The presented refrigerator is a general purpose device which allows for a broad variety of different applications. Our focus lies on particle physics developments and experiments \cite{Taqqu2006,Bao2014,URREGOBLANCO2007}, neutron particle physics and scattering \cite{Piegsa2009c,vdBrandt2009,Piegsa2013b,Wichmann2016}, ultracold neutron production \cite{SchmidtWellenburg2006799,Zimmer2011,Piegsa2014b,Leung2015}, and dynamic nuclear polarisation applications in bio-medical research \cite{NBM:NBM1682,CMR:CMR20099}.

\section{Cryostat system}
\begin{figure}
	\centering
		\includegraphics[width=0.40\textwidth]{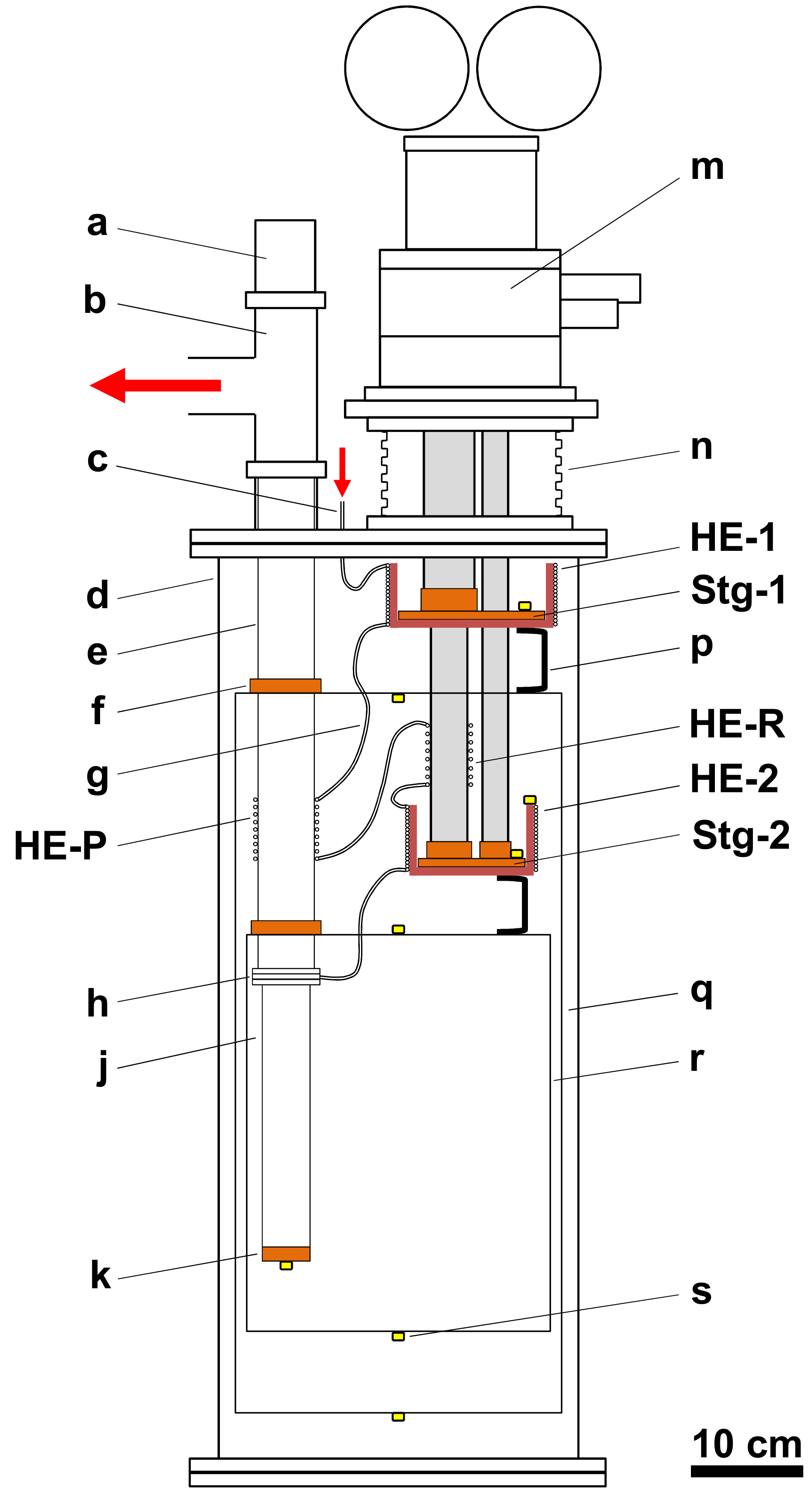}
	\caption{Schematic drawing of the cryostat without cryostat-insert: (a)~flange for the cryostat-insert and top-loading sample-stick, (b)~DN 50 KF vacuum T-piece connected to pumping set, (c)~helium gas inlet, (d)~DN 320 ISO-K vacuum vessel with a height of 800~mm, (e)~stainless steel pumping tube (49~mm i.d., 0.5~mm wall thickness), (f)~copper clamp serving as thermal anchor between radiation shield and pumping tube, (g)~helium gas capillary, (h)~conus flange leading to a cold needle valve mounted on the insert, (j)~lower part of the stainless steel pumping tube (42~mm i.d., 0.5~mm wall thickness), (k)~bottom of the 1 K pot with an attached resistive heater of 120~$\Omega$, (m)~pulse tube cryocooler, (n)~DN 160 ISO-K bellow, (p)~copper braid straps for thermal anchoring, (q)~cylindrical outer radiation shield with a height of 640~mm and an outer diameter of 290~mm, (r)~cylindrical inner radiation shield with a height of 350~mm and an outer diameter of 270~mm, (s)~calibrated thermometers, (HE-1, HE-2, HE-P, HE-R)~gas heat exchangers attached to the first and second stage of the cryocooler, pumping tube, and regenerator tube, respectively, and (Stg-1, Stg-2)~first and second stage of the cryocooler. }
	\label{fig:Fig1_Cryo_Scheme}
\end{figure}

\begin{figure}
	\centering
		\includegraphics[width=0.30\textwidth]{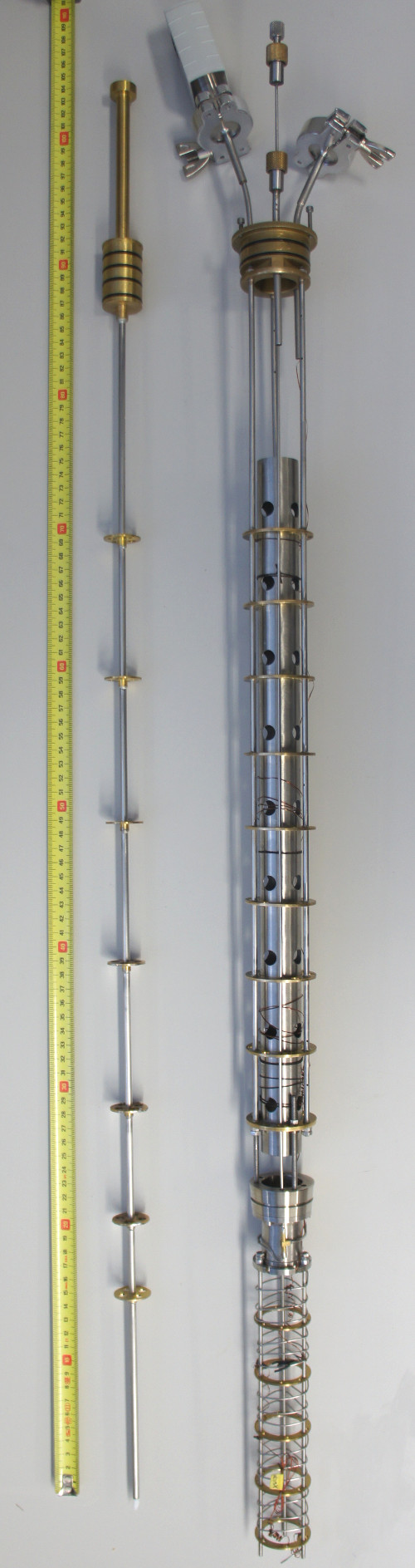}
	\caption{Picture of the sample-stick (left) and the Roubeau-type $^4$He cryostat-insert (right).}
	\label{fig:Fig2_Insert}
\end{figure}

\begin{figure*}
	\centering
		\subfigure[(a)]{\includegraphics[width=0.33\textwidth]{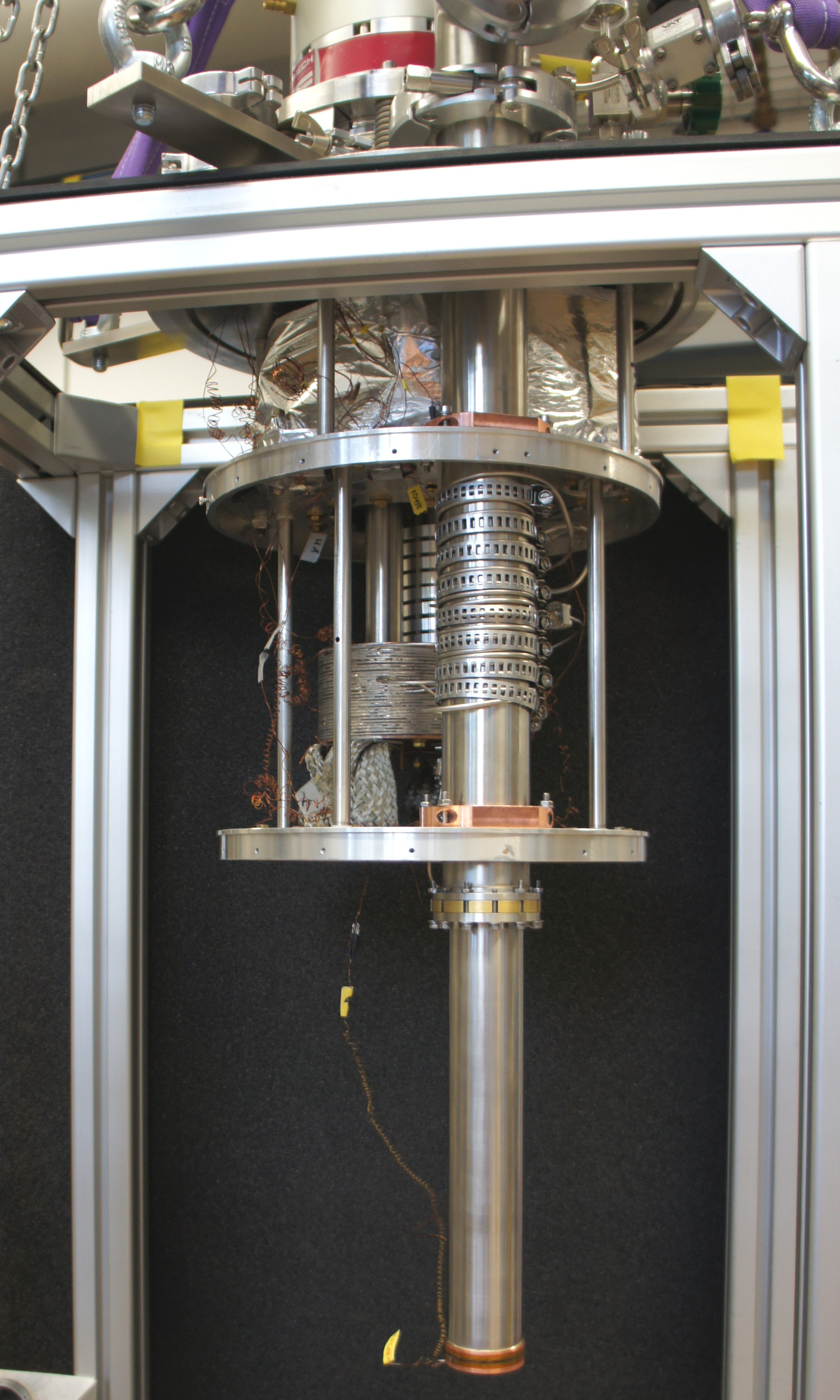}}
		\hspace{1 cm}
		\subfigure[(b)]{\includegraphics[width=0.33\textwidth]{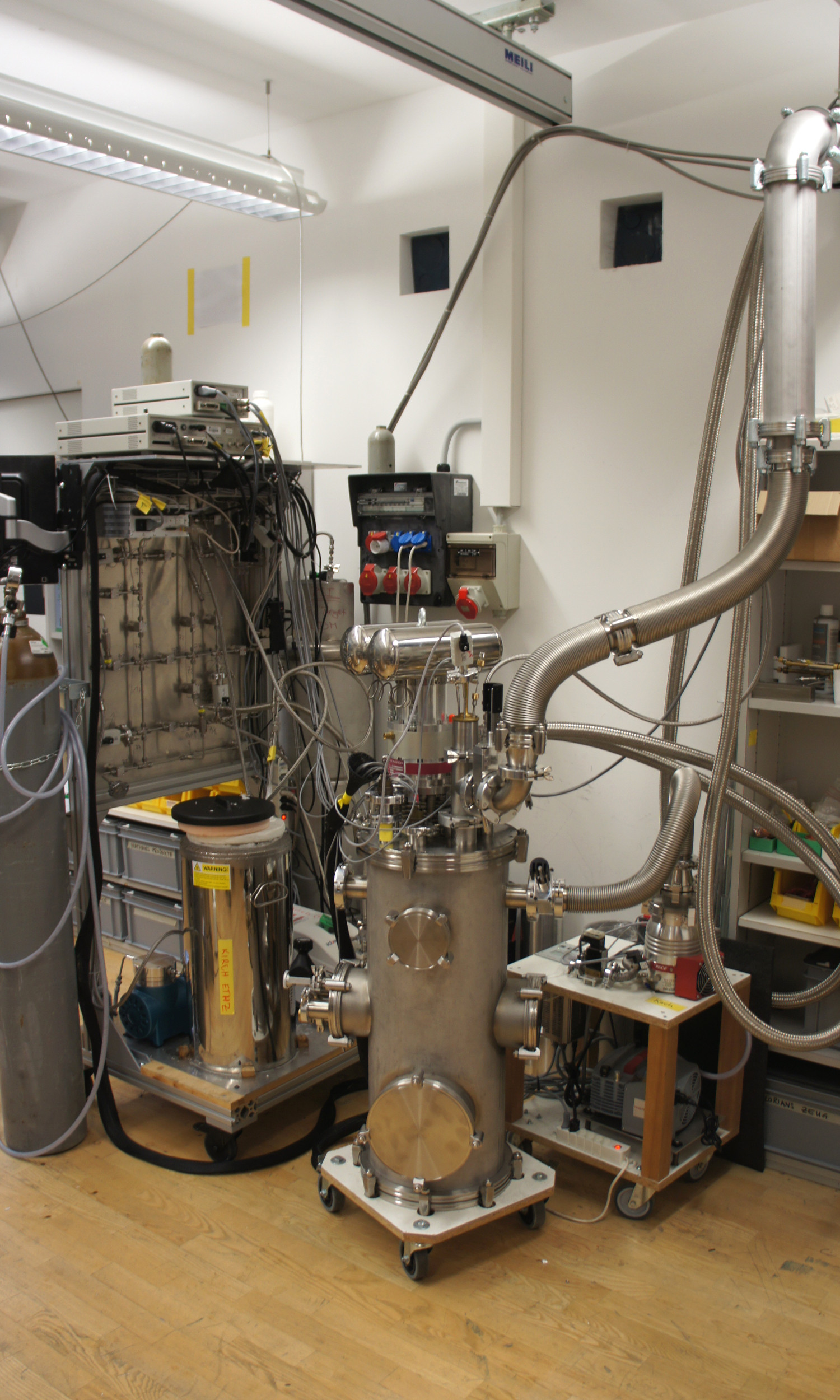}}\\
		\vspace{0.5cm}
		\subfigure[(c)]{\includegraphics[width=0.33\textwidth]{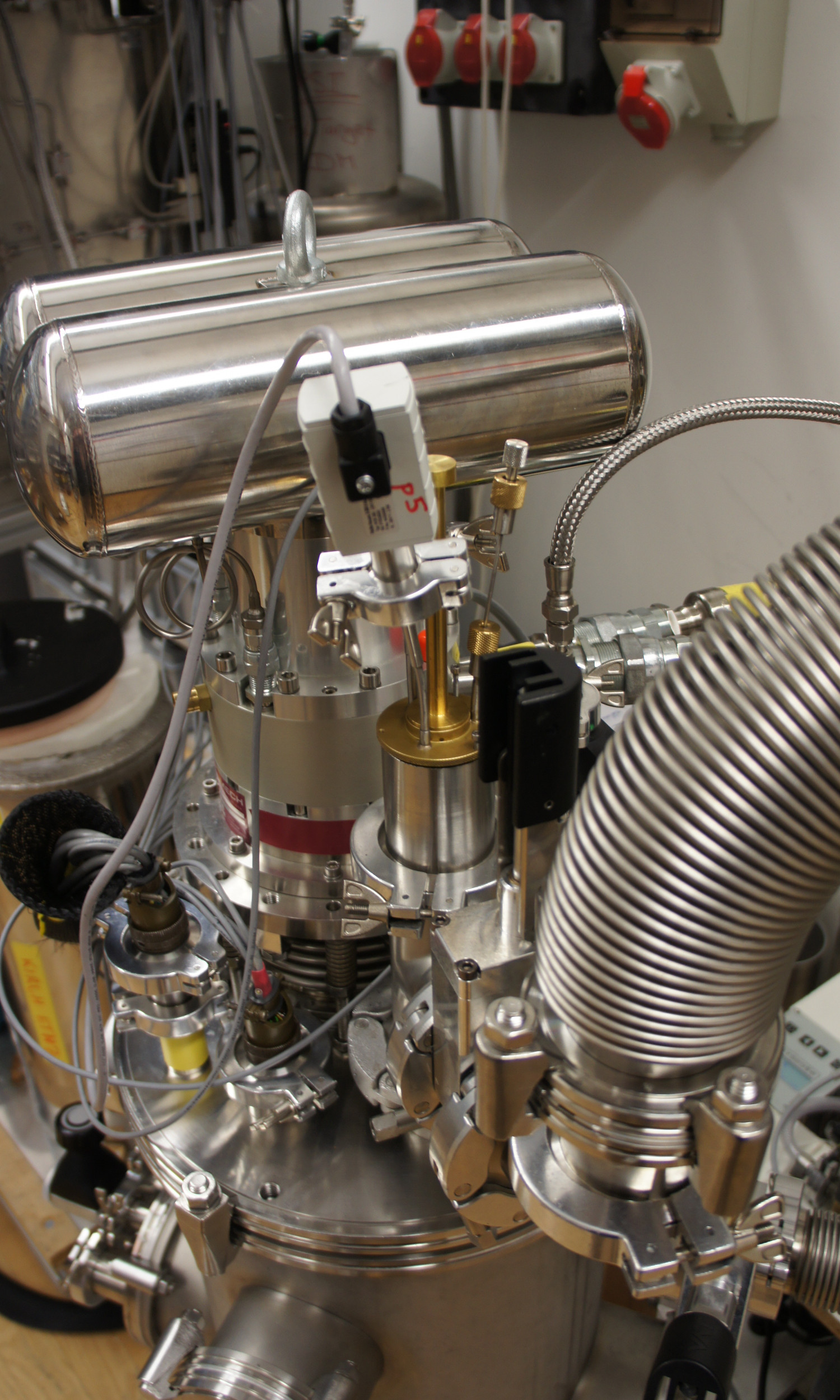}}
		\hspace{1 cm}
		\subfigure[(d)]{\includegraphics[width=0.33\textwidth]{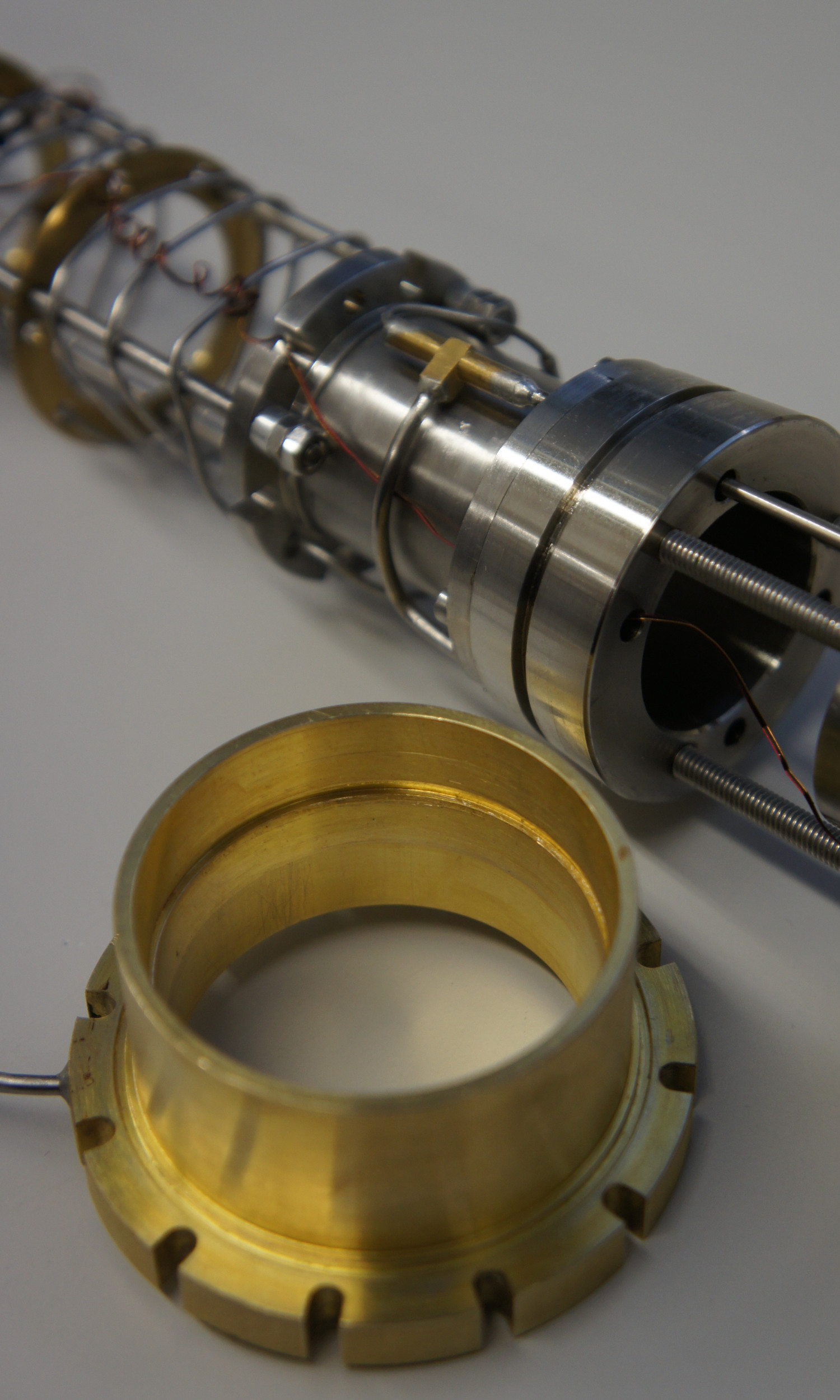}}
	\caption{(a)~Opened cryostat: the pumping tube is thermally anchored to the lids of the two thermal radiation shields with copper clamps. The rest of the shields are removed. The capillary of the HE-P is attached to the pumping tube with eight hose clamps. In the background the HE-2 is visible with the capillary brazed to the copper cup/cylinder. (b)~Cryostat setup in the laboratories of ETH Z\"urich with the gas-handling system (left), a turbo pump for the isolation vacuum (right), and tubes to the pumping set in the adjacent room. (c)~Top-view of the cryostat. (d)~Detailed view of the brass conus flange, its stainless steel counterpart, and the cold needle valve.}
	\label{fig:Fig3_Cryo}
\end{figure*}

A detailed scheme and description of the cryostat is presented in Fig.\ \ref{fig:Fig1_Cryo_Scheme}. A commercial PTC\footnote{\emph{Cryomech}, PT415.} with a nominal cooling power of 1.5~W at 4.2~K and a pumping tube are both mounted into a DN 320 ISO-K tube of length 800~mm which serves as an isolation vacuum vessel. 
The pumping tube is made from thin walled stainless steel pipes and hosts the cryostat-insert. On its bottom end it is terminated with a brazed copper disc (1 K pot). The disc can be heated using a resistive film heater. Attached to the top/warm end of the pumping tube is a DN 50 KF T-piece which itself is connected via a DN 63 ISO-K bellow and a DN 100 ISO-K tube (total length of approximately 4~m) to a hermetic pumping set. The set consists of two roots pumps in series, with nominal pumping speeds of 1000~m$^3$/h and 300~m$^3$/h, backed by a 35~m$^3$/h scroll pump.
The transmission of vibrations from the PTC to the residual apparatus are suppressed by decoupling them by means of a commercial DN 160 ISO-K bellow\footnote{\emph{Metallicflex}, Germany.} with spring suspension and by attaching all thermal anchors using flexible copper braid straps, e.g.\ two thermal radiation shields are attached to the corresponding two stages of the PTC. The shields each consist of a top aluminum lid with the thermal connection, a bottom aluminum cap, and a  2~mm thick aluminum rolled cylinder covered with a 10-layers super-insulation foil.\footnote{\emph{RUAG}, Coolcat 2-NW.} The pumping tube is thermally anchored to the lids of the shields with rigid copper clamp rings.\\
The cryostat employs a closed cooling cycle using helium gas - currently we are using only $^4$He. The gas is continuously filtered by passing it through a charcoal cold trap to prevent blocking of capillaries by gas impurities before it is introduced into the cryostat via a feed-through at the large top vacuum flange. 
The corresponding coldtrap is immersed in a liquid nitrogen bath located external to the cryostat.\footnote{Such a coldtrap can easily be integrated into the cryostat setup to achieve an entirely dry system. However, during the development phase an external coldtrap is more comfortable.}
The helium gas is precooled and eventually liquefied by means of four successive heat exchangers (HE) attached to the first and second stage of the cryocooler (HE-1 and HE-2), to the regenerator between the two stages (HE-R), and to the pumping tube (HE-P). The actual sequence of the gas passing through the heat exchangers (HE-1, HE-P, HE-R, and HE-2) is indicated in Fig.\ \ref{fig:Fig1_Cryo_Scheme}. HE-1 and HE-2 both consist of copper cups with stainless steel gas capillaries brazed onto them. The cups are directly screwed to the stages of the PTC. The capillaries each have a length of approximately 10~m and the dimensions $\diameter 2 \times 3$~mm for HE-1 and $\diameter 1 \times 2$~mm for HE-2.\footnote{Here $\diameter 1 \times 2$~mm refers to a capillary with an inner diameter of 1~mm and an outer diameter of 2~mm.}
The capillary of HE-P ($\diameter 2 \times 3$~mm, length about 3.5~m) is wound around the pumping tube and attached using hose clamps.
The capillary of HE-R ($\diameter 1 \times 2$~mm, length about 1.5~m) is meandering on the surface of the regenerator of the PTC. It is attached with aluminum clamps.
For development and upgrade flexibility, the individual capillaries of the heat exchangers are connected using small indium sealed flanges. This allows dismounting and replacing of all heat exchangers without soldering, brazing, or other severe mechanical interventions. 
The liquefied helium is transferred into the pumping tube via an inlet and a dedicated cold seal. The sealing consists of a brass conus mounted inside the pumping tube and its stainless steel counterpart on the cryostat-insert. The two parts are precisely machined such that they exactly match each other and form a helium tight (warm and cold) connection. The liquid helium is inserted through a small groove in the conus and a short capillary to a cold needle valve on the insert, see Fig.\ \ref{fig:Fig3_Cryo}d. The valve allows adjusting the flow of the helium which is finally condensed into the 1 K pot. In addition, the pumped vapor of the helium bath causes a counter flow cooling of HE-P and the thermal radiation shields.\\
In Fig.\ \ref{fig:Fig2_Insert} the $^4$He cryostat-insert and the corresponding sample-stick are depicted. They each consist of a frame structure made from stainless steel tubes and several different types of brass baffles which prevent a direct line of sight between the 1 K pot and room temperature.
The sample-stick with an O-ring sealed brass head (29~mm o.d.) allows for a quick and frequent sample loading/exchange directly into the cold 1 K pot during operation. It is also possible to rotate the stick along the axis of the pumping tube and, thus, a sample can be oriented with respect to an external magnetic field or particle beam.
In contrast, the cryostat-insert usually remains in the pumping tube and is only removed after a warm-up of the entire system. The brass head of the insert (49~mm o.d.) with a central opening for the sample-stick also has feed-throughs for measuring the vapor pressure in the pumping tube and the temperature of the helium bath.
Instead of loading samples directly into the helium bath, the refrigerator can also be used to cool down objects (e.g.\ gas cells, electronic devices, particle detectors etc.) by placing them into the isolation vacuum and attaching them to the copper disc of the 1 K pot. In Fig.\ \ref{fig:Fig3_Cryo} several images of the apparatus are presented.

\begin{figure}
	\centering
	\includegraphics[width=0.45\textwidth]{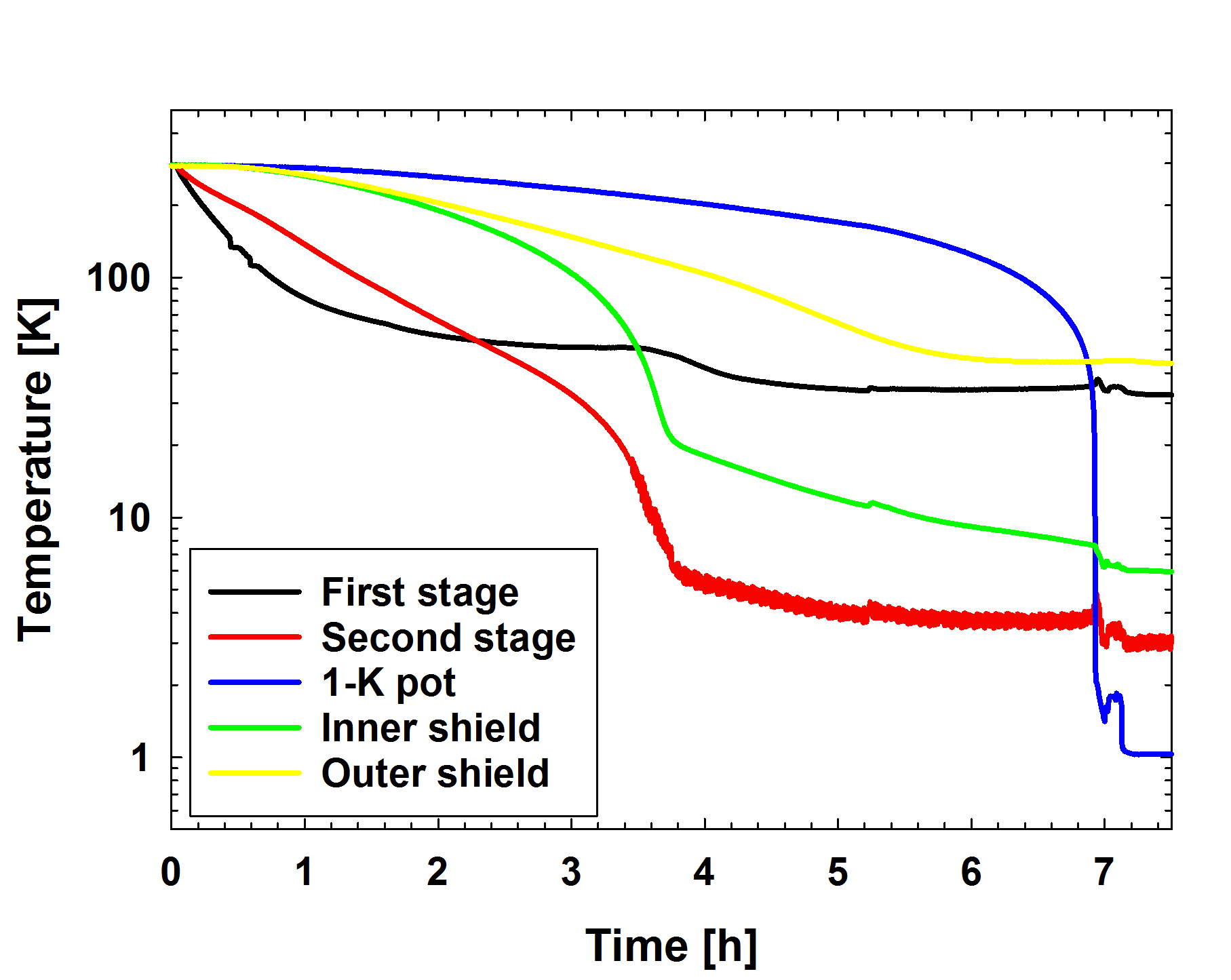}
	\caption{Measured temperatures during cool-down of the cryostat starting from room temperature. The temperatures of the two stages of the PTC settle at approximately 30~K and 3~K, respectively. The temperatures of the outer and inner thermal radiation shields are measured on their bottom aluminum caps. They settle at about 45~K and 6~K. For all temperature measurements calibrated Cernox thermometers with a typical sensor accuracy of $\pm 5$~mK are used.}
	\label{fig:Fig4_Cooldown}
\end{figure}

\begin{figure}
	\centering
		\subfigure[(a)]{\includegraphics[width=0.45\textwidth]{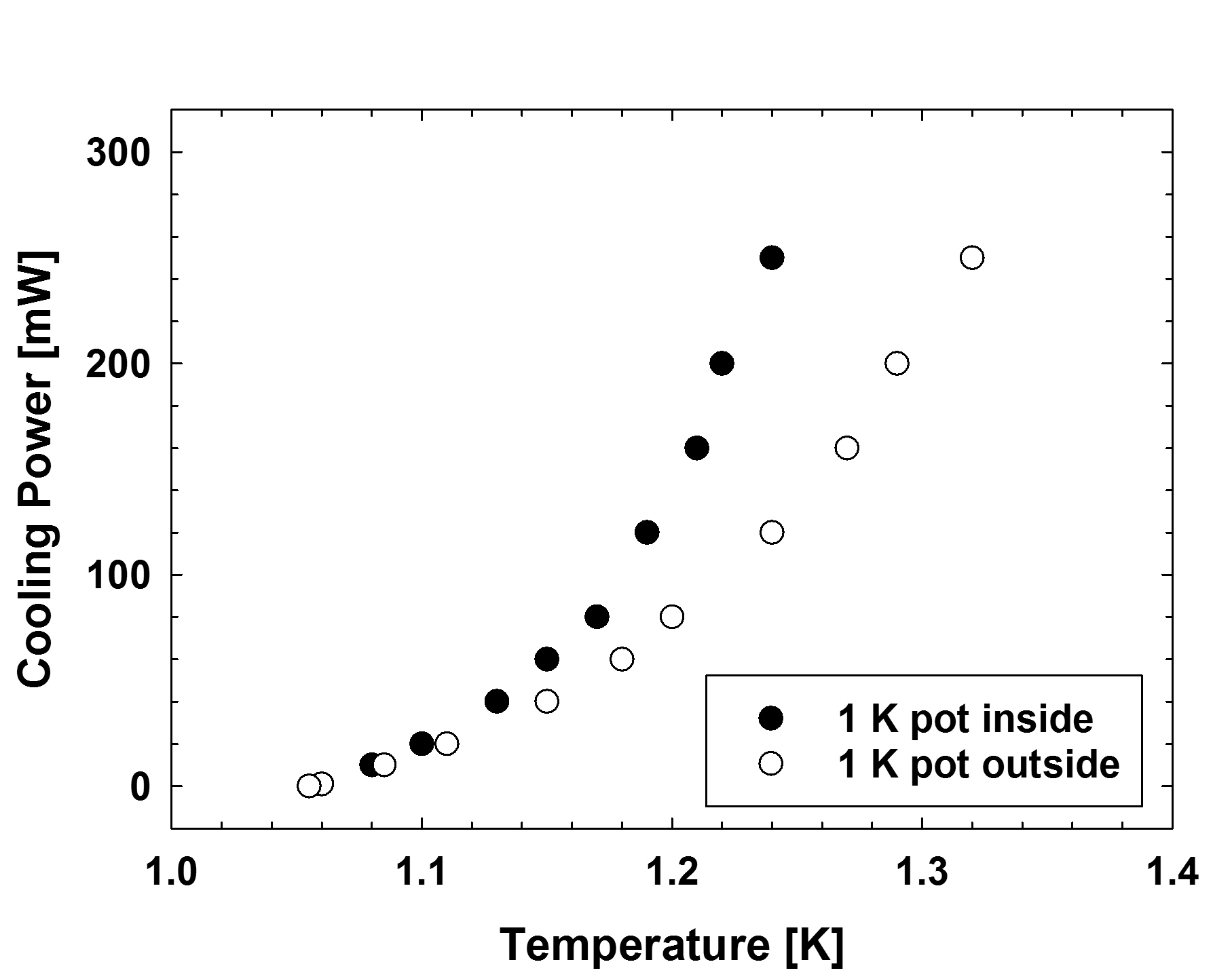}}
	  \subfigure[(b)]{\includegraphics[width=0.45\textwidth]{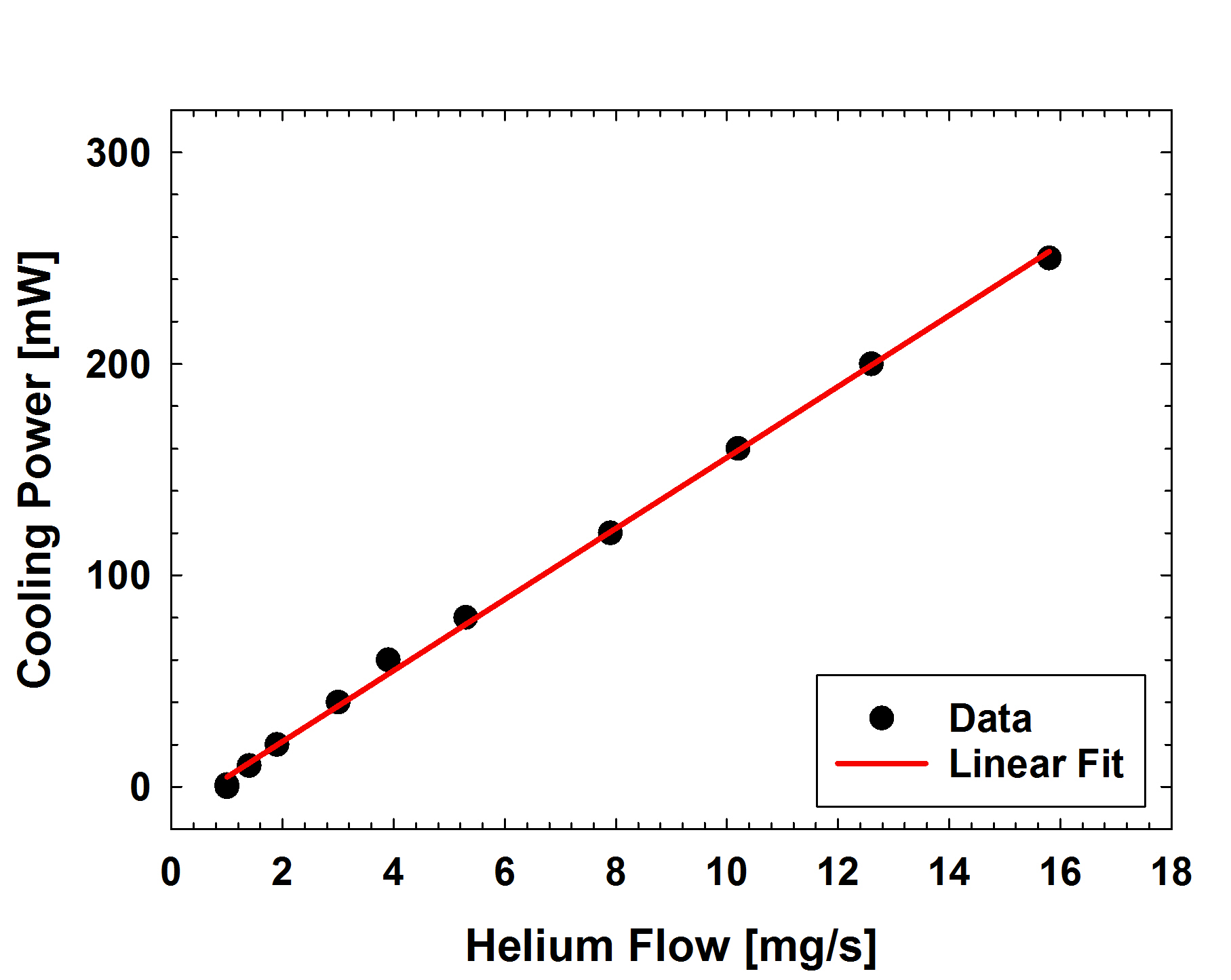}}
		\subfigure[(c)]{\includegraphics[width=0.45\textwidth]{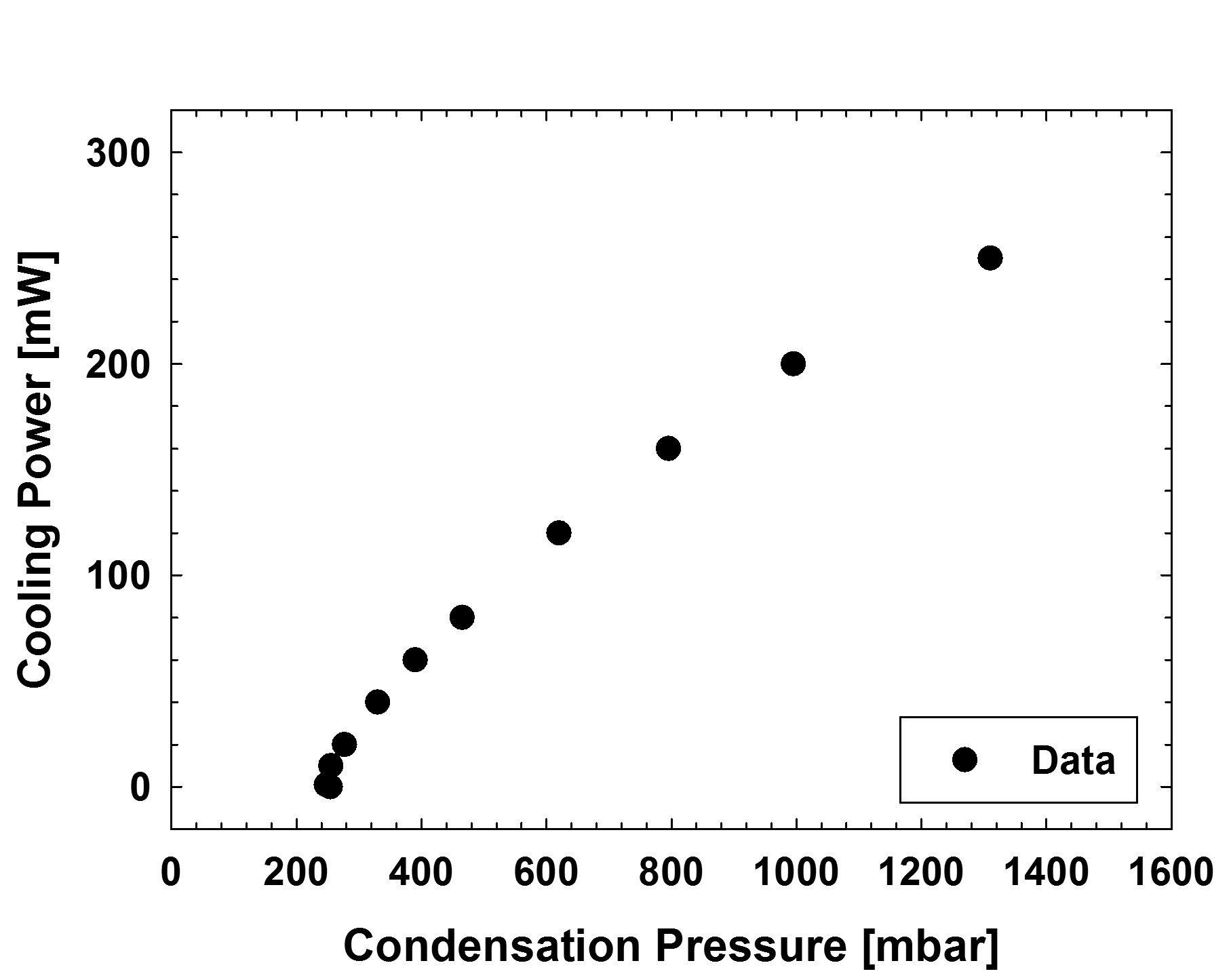}}
	\caption{Cooling power of the cryostat as a function of (a)~1 K pot temperature, (b)~helium flow, and (c)~condensation pressure. The 1 K pot temperature is measured with one thermometer immersed in the helium bath (inside) and one attached to the copper disc in the isolation vacuum (outside). The cooling power increases linearly with the helium flow. }
	\label{fig:Fig5_Heating}
\end{figure}

\section{Operation and performance of the cryostat}

The entire cooling procedure of the refrigerator setup down to the base temperature of about 1.0~K takes approximately seven hours, as shown in Fig.\ \ref{fig:Fig4_Cooldown}. The isolation vacuum is evacuated using a turbo pump. After it reaches a value below $10^{-4}$~mbar the PTC can be started. During the cool-down the isolation vacuum pressure successively drops until it reaches about $10^{-7}$~mbar. A small amount of helium gas is continuously circulated through the closed-cycle as a contact gas. To achieve gas flows of a few mg/s the inlet gas pressure is increased to up to 2.5~bar by means of a small hermetic compressor unit.\footnote{\emph{KNF}, membrane compressor N143AN.12E.} 
At that time only the scroll pump is used for circulation and the cold needle valve is fully opened. To establish a stable flow helium can be added from an external storage gas tank about once every two hours. The tank has a volume of 100~liters and is initially filled with up to 1~bar of helium. After the 1 K pot reaches a temperature of about 2~K, the compressor unit is stopped and normal circulation is started. Now additional helium is condensed from the gas tank to fill the 1 K pot (typical total helium amount of 5~g). Finally, the roots pumps are started, the needle valve is adjusted, and the cryostat temperature quickly drops to its base temperature.\\
The cooling power of the cryostat is measured using the resistive film heater attached to the outside of the 1 K pot. The results are summarized in Fig.\ \ref{fig:Fig5_Heating}. When the heater is switched off temperatures of just above 1.0~K are achieved with a condensation pressure on the helium gas inlet of approximately 250~mbar and a helium flow of 1~mg/s.
With increased heating the condensation pressure and the helium flow rise.
At a temperature of 1.24~K of the helium bath a continuous cooling power of 250~mW is obtained.
In this state a condensation pressure of 1300~mbar and helium flow of about 15.8~mg/s is reached. 
Not plotted is the vapor pressure over the helium bath measured with the pressure gauge on the top of the cryostat insert, which increases from about 0.1~mbar (at 0~mW) to 0.6~mbar (at 250~mW). 
A linear fit to the helium flow data in Fig.\ \ref{fig:Fig5_Heating}b results in a value for the slope of $(16.8 \pm 0.2)$~J/g. This is close to the value of the latent heat of evaporation of $^4$He of about 20~J/g and thus indicates a high efficiency of the cryostat.
\begin{figure}
	\centering
		\includegraphics[width=0.45\textwidth]{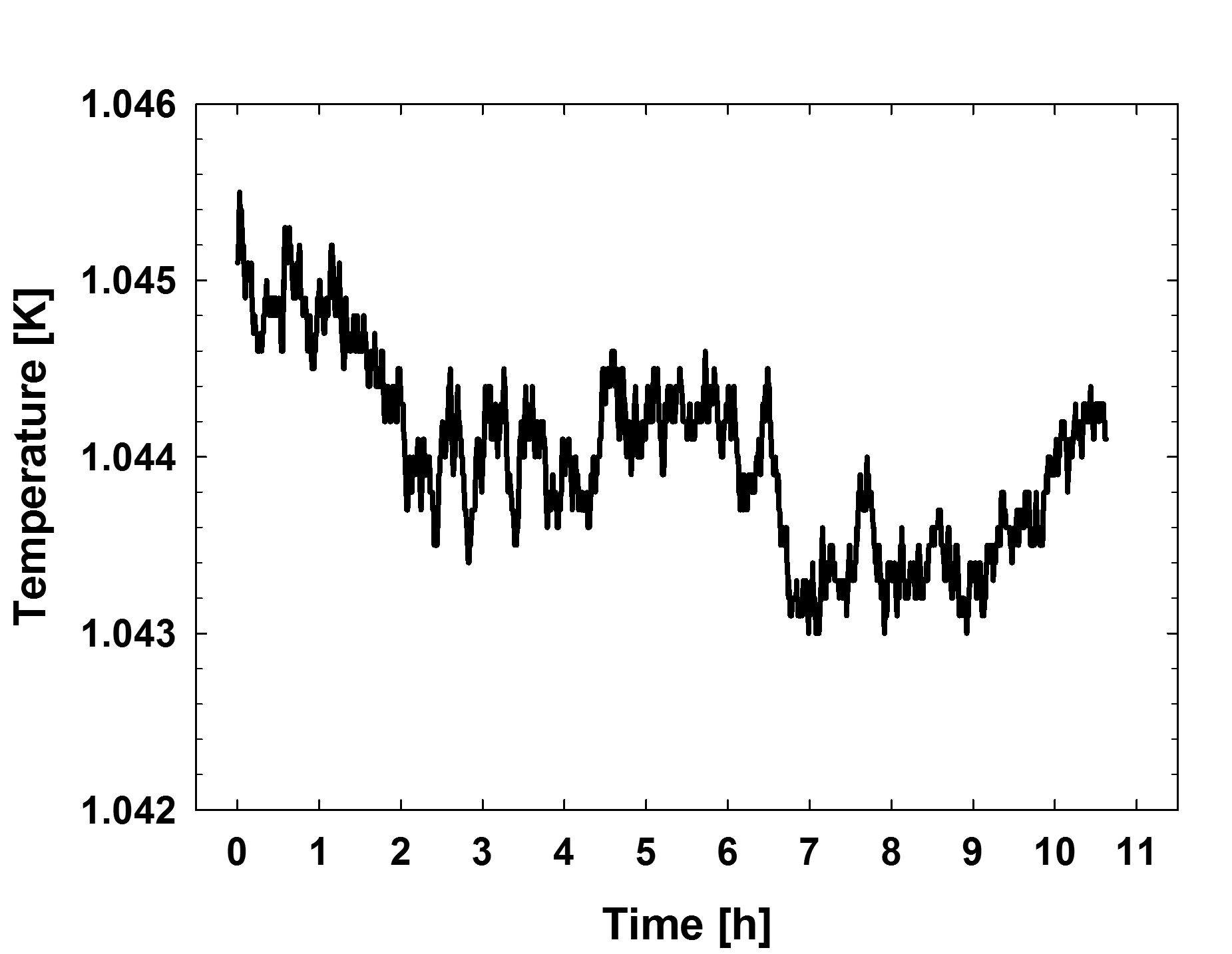}
	\caption{Temperature stability of the 1 K pot (outside) during an over night measurement. }
	\label{fig:Fig6_stab}
\end{figure}\\
The refrigerator runs very reliably and can be operated for several days continuously without close surveillance. In Fig.\ \ref{fig:Fig6_stab} an exemplary temperature stability measurement is presented. Over a period of more than 10~hours no variations of the 1 K pot temperature of larger than $\pm 1$~mK are observed.\\
Finally, to load a (new) sample the liquefied helium has to be pumped back into the storage tank. This procedure can be accelerated by employing the resistive film heater. Then the sample-stick can be removed and reloaded through the central opening of the cryostat-insert. To prevent air from condensing in the cold cryostat, helium gas is flushed into the pumping tube via a separate inlet. If the sample-stick is precooled in liquid nitrogen before reloading, the 1 K pot reaches its base temperature again within less than 75~min.


\section{Conclusions}

We presented a high-power closed-cycle cryostat with top-loading sample exchange. The apparatus is versatile, reliable, and user-friendly. In can be used for a broad range of applications. 
Possible upgrades and future developments could consist in a more powerful PTC, optimized heat exchangers, and a reduction of masses (e.g.\ copper cups of HE-1 and HE-2) to increase the cool-down speed etc.
The low condensation pressure of 250~mbar for $^4$He indicates that $^3$He could be directly condensed only by means of the second stage of the cryocooler. Hence, we are currently investigating the possibility to operate a $^3$He/$^4$He dilution cryostat-insert in the same apparatus.

\section{Acknowledgments}

We gratefully acknowledge the excellent technical support by Paul Schurter from PSI and the D-PHYS physics department workshop at ETH Z\"urich.
Moreover, we appreciate many fruitful discussions with Marek Bartkowiak, Patrick Hautle, and Ton Konter. This research was financed in part by the Swiss National Science Foundation under grant numbers 200020-159754 and 200020-172639.





\bibliographystyle{elsarticle-num2}
\bibliography{piegsa-bibfile}


\end{document}